\documentclass{sig-alternate}

\usepackage[
  pass,
]{geometry}

\usepackage[export]{adjustbox}
\usepackage[hang]{subfigure}

\usepackage{cite}
\usepackage{hyperref}
\usepackage{wrapfig}

\begin{document}
%
\conferenceinfo{FSE'14}{, November 16--22, 2014, Hong Kong, China}
\CopyrightYear{2014}
\crdata{978-1-4503-3056-5/14/11}

\title{Linking Sketches and Diagrams to Source Code Artifacts}
%
%
%
%
%

\numberofauthors{1} 
%
\author{
%
%
\alignauthor
Sebastian Baltes, Peter Schmitz, and Stephan Diehl \\
       \affaddr{Computer Science}\\
       \affaddr{University of Trier}\\
       \affaddr{Trier, Germany}\\
       \email{\{s.baltes,diehl\}@uni-trier.de}
}

\maketitle
\begin{abstract}
Recent studies have shown that sketches and diagrams play an important role in the daily work of software developers.
If these visual artifacts are archived, they are often detached from the source code they document, because there is no adequate tool support to assist developers in capturing, archiving, and retrieving sketches related to certain source code artifacts.
This paper presents \emph{SketchLink}, a tool that aims at increasing the value of sketches and diagrams created during software development by supporting developers in these tasks.
Our prototype implementation provides a web application that employs the camera of smartphones and tablets to capture analog sketches, but can also be used on desktop computers to upload, for instance, computer-generated diagrams.
We also implemented a plugin for a Java IDE that embeds the links in Javadoc comments and visualizes them in situ in the source code editor as graphical icons.
\end{abstract}

\begin{figure}[h!]
\noindent More information: \url{http://st.uni-trier.de/sketchlink}
\vspace{-\baselineskip}
\end{figure}




\begin{figure*}[t!]
\centering
\subfigure[Whiteboard sketching]{\includegraphics[width=0.3\textwidth, frame]{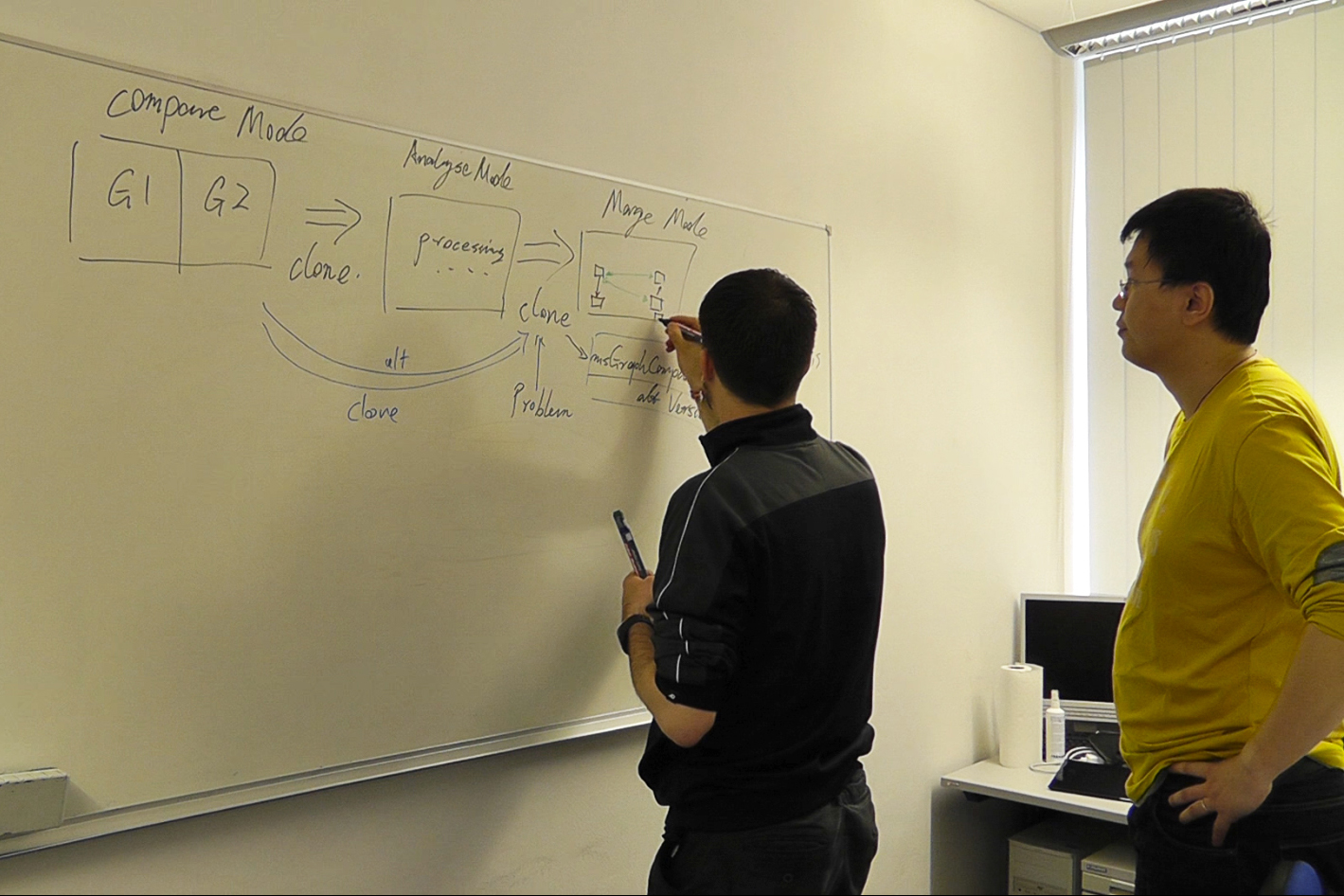}}
\hfill
\subfigure[Capturing the sketch]{\includegraphics[width=0.3\textwidth, frame]{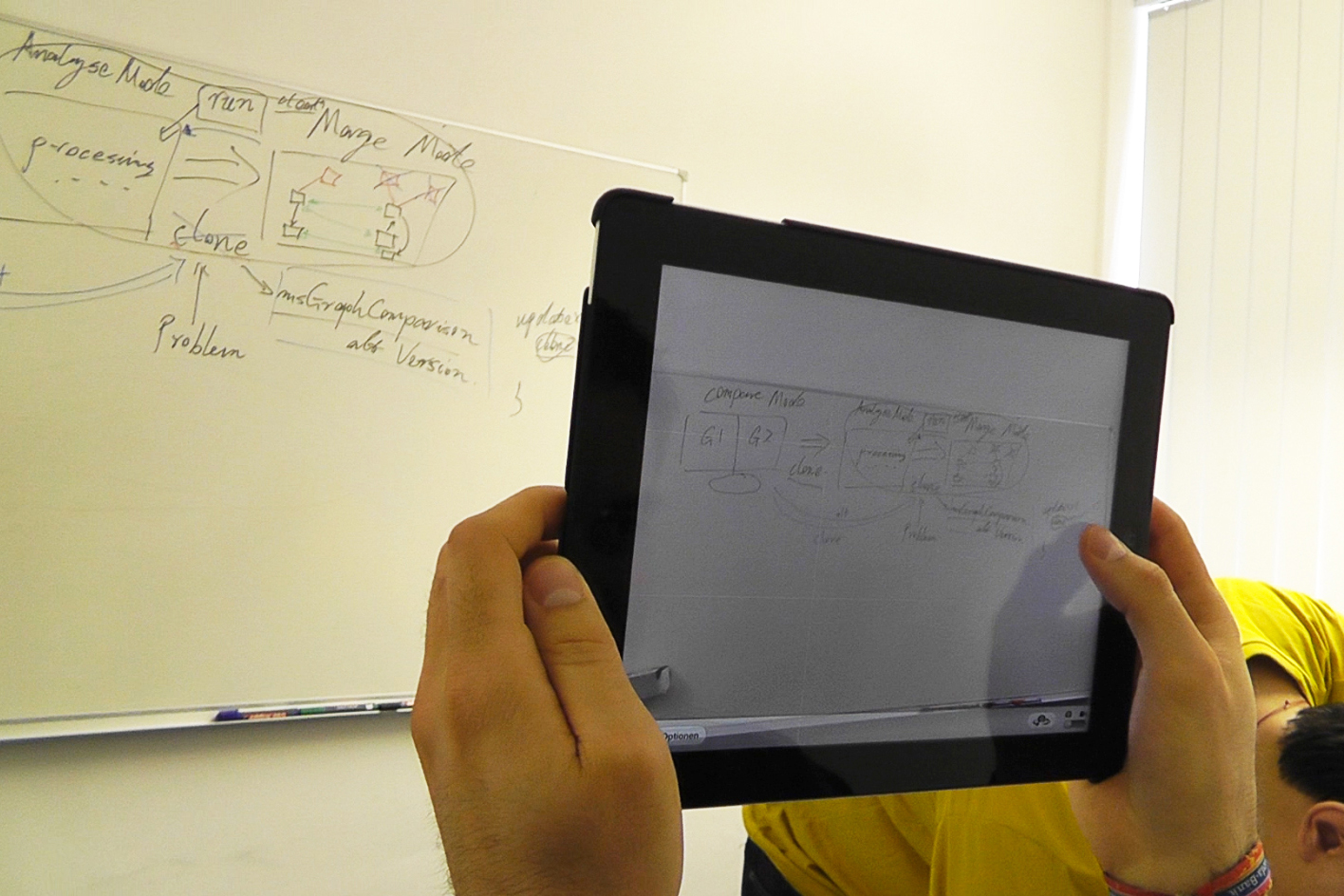}}
\hfill
\subfigure[Selecting and linking areas]{\includegraphics[width=0.3\textwidth, frame]{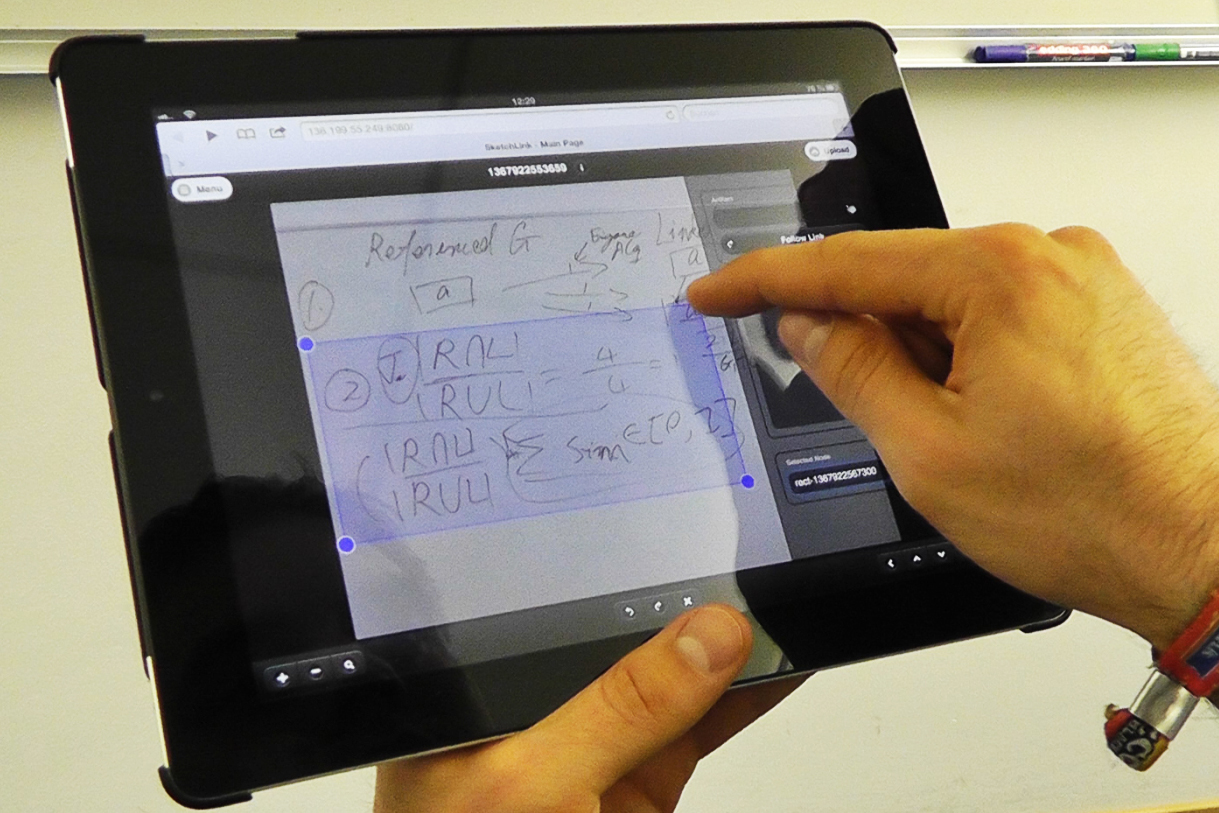}}
\caption{Exemplary usage of SketchLink for whiteboard sketching}
\label{fig:workflow}
\end{figure*}

\section{Introduction}

Sketches and diagrams play an important role in the daily work of software developers~\cite{Dekel07, Cherubini07, Walny11-1, Baltes14}.
Most of these visual artifacts do not follow formal conventions like the \emph{Unified Modeling Language} (UML), but have an informal, ad-hoc nature~\cite{Cherubini07, Petre13, Dekel07, Baltes14, Gorschek14}.
They may contain different views, levels of abstraction, formal and informal notations, pictures, or generated parts~\cite{Hoek14, Dekel07, Cherubini07, Walny11-2}.
Developers create sketches mainly to understand, to design, and to communicate~\cite{Cherubini07, Baltes14}.
Media used for sketch creation include not only whiteboards and scrap paper, but also software tools like Photoshop and PowerPoint~\cite{Walny11-1, Cherubini07, Myers08, Gorschek14}.
Sketches and diagrams are important because they depict parts of the mental model developers build to understand a software project~\cite{LaToza06}.
Understanding source code is one of the most important problems developers face on a daily basis \cite{Cherubini07, LaToza06, Singer97, Ko06}.
However, this task is often complicated by documentation that is frequently poorly written and out of date~\cite{Forward02, Lethbridge03}.
Sketches and diagrams, whether formal or informal, can fill in this gap and serve as a supplement to conventional documentation like source code comments.
To this end, tool support is needed to assist developers in archiving and retrieving sketches and diagrams related to certain source code artifacts.

\section{Related Work}

In the past, many tools have been proposed that aim at supporting developers' sketching activities.
Some of them force users to employ special devices like electronic whiteboards~\cite{Damm00, Chen03, Mangano10} or digital pens and paper~\cite{Dachselt08, Brandl08}.
These tools often focus on UML as they try to convert sketches into formal UML diagrams~\cite{Damm00, Chen03, Dachselt08, Hammond06}.
Branham et al.\ proposed a tool to automatically capture whiteboard drawings using a networked camera~\cite{Branham10}. This approach leads to a large number of archived sketches, which are not likely to be used in the future. Again, special hardware is needed for capturing the drawings.
Furthermore, tools exist that allow to create sketches directly in the source code editor of IDEs~\cite{Brandl08, Hammond06}.
This approach is also of limited use, because, on the one hand, sketching may happen in design meetings with other stakeholders where IDEs are not used and source code is not immediately created.
On the other hand, sketches and diagrams may provide a high-level understanding of the project architecture \cite{Lee08} and may thus be linked to different artifacts in different source code files.
These use cases are difficult to support if sketches are created directly in the source code editor and are attached to a single source code file.

In our opinion, existing tools do not adequately consider developers' needs.
In a recent study, Petre observed that software developers ``will not adopt tools and ideologies at odds with their considered practice''~\cite{Petre13}.
Walny et al.\ note that a tool integrating sketches into the software development workflow must support a broad range of work styles~\cite{Walny11-1}, which most of the above mentioned tools do not achieve.

\section{Motivation}

In a large study with 394 participants \cite{Baltes14}, we investigated the use of sketches and diagrams in software engineering practice to validate our motivation for building a new tool.
In this section, we briefly summarize the most important findings that are relevant for the design of this tool.
As expected, the majority of sketches and diagrams from the study were informal and most of them were drawn on analog media like paper or whiteboards. 
The most common purposes for creating sketches were related to designing, explaining, or understanding. 
One third of them had an estimated lifespan of one day or less, one third of up to one month, and another third of more than one month.
The majority of sketches were archived, most of them digitally. Many sketches were kept because they document or visualize parts of the implementation or assist its understanding.
The high number of archived sketches lead to the assumption that developers are willing to keep their visual artifacts. However, they also named technical issues, e.g., that there is no good technique to keep sketches together with source code.
Regarding the relation to source code, we found out that sketches and diagrams were rarely related to certain attributes or statements, but rather to methods, classes, packages, or projects (or, depending on the programming language, other artifacts with the same levels of abstraction).
About half of the sketches and diagrams from our study were rated as helpful to understand the related source code artifact(s) in the future, which supports our goal to use sketches as a supplement to conventional documentation.

\section{Our Approach}

With the results from our study in mind, our main goal was to create a tool that would enable developers to easily capture and annotate the sketches and diagrams they create to link them afterwards to the related source code artifacts.
The sketches could then be used to understand the related code and to navigate to the linked artifacts, enabling developers to explore relations depicted in the linked sketches.
The tool should integrate with heterogenous workflows and should not be restricted to a certain visual convention or a special medium for creating sketches.
Finding relevant documentation in external systems is a task that developers generally regard as challenging, time consuming, and not always worth its effort, because even an elaborate search does not guarantee to produce helpful content \cite{Lethbridge03, LaToza06}.
Thus, the links should be visualized in situ in the source code editor, e.g., using color coding, highlighting, or graphical icons, but should not distract the developer.
This allows developers to quickly access relevant sketches.
In order to provide flexible means for capturing analog drawings, we decided to focus on mobile devices like smartphones and tablets for capturing, annotating, and linking sketches.
Since such devices are prevalent nowadays, they are available in almost every situation (see Figure \ref{fig:workflow} for an exemplary workflow).
However, it should also be possible to upload, for instance, computer-generated diagrams from conventional desktop computers.

\begin{figure*}[t!]
\centering
\subfigure[Floating mode]{\includegraphics[width=0.36\textwidth, frame]{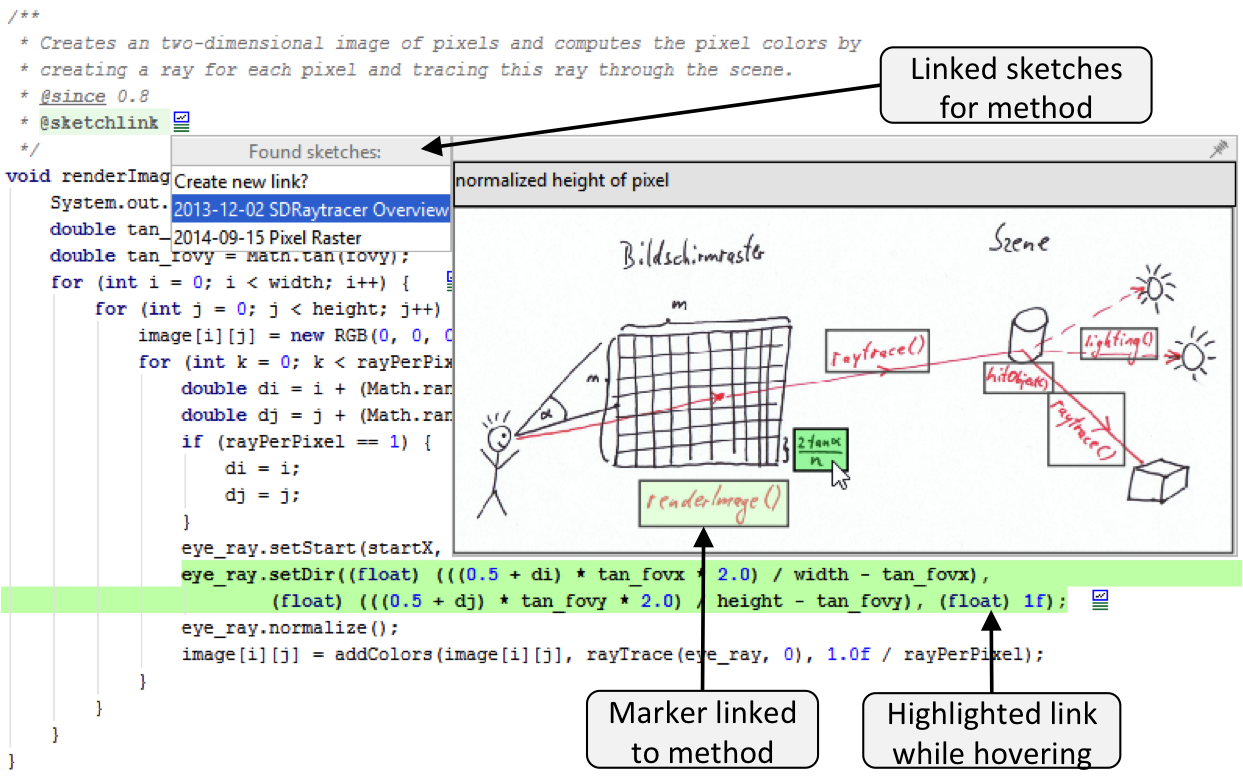}}
\hfill
\subfigure[Docked mode]{\includegraphics[width=0.63\textwidth, frame]{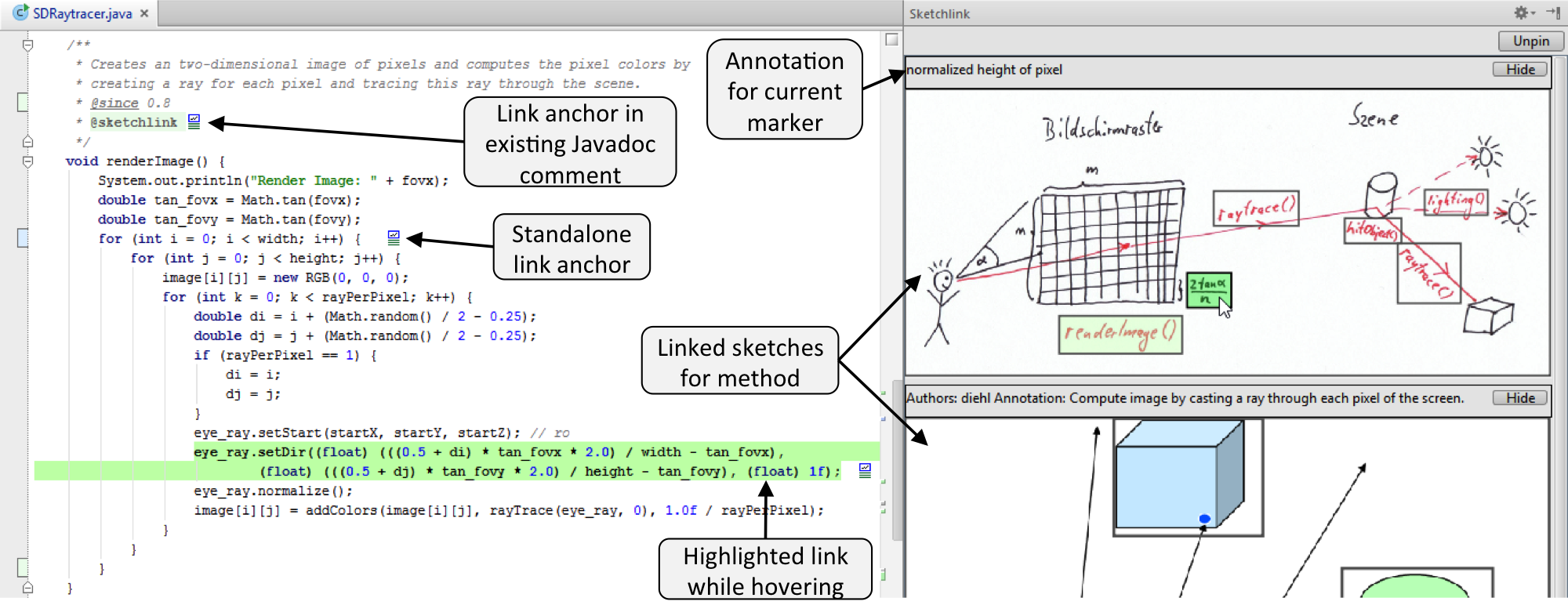}}
\caption{SketchLink plugin for IntelliJ IDEA}
\label{fig:plugin}
\end{figure*}

\begin{figure}[b!]
\centering
\includegraphics[width=\columnwidth]{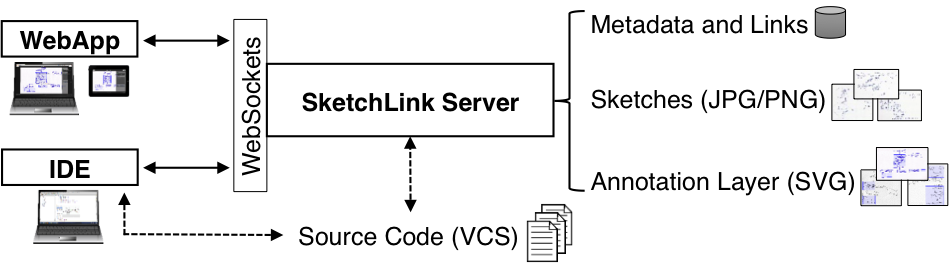}
\caption{SketchLink architecture}
\label{fig:architecture}
\end{figure}

\section{Prototype}

Our prototype named \emph{SketchLink} consists of a server, a web application, and an IDE plugin (see Figure \ref{fig:architecture}).
The server stores the sketch images, metadata, and the links, providing a WebSocket interface for updating and retrieving this information.
The web application runs in both desktop and mobile browsers and can be used to upload, annotate, and link sketches.
It requests information about available source code artifacts from the server, which has access to the version control system. 
The IDE plugin, which we implemented for the Java IDE \emph{IntelliJ IDEA}, visualizes the links in the editor and can be used to create link anchors in the source code.
Furthermore, the plugin enables the web application to scroll the editor view to a linked source code artifact to navigate through source code using a linked sketch or diagram.

\begin{figure}[b]
\centering
\vspace{-\baselineskip}
\includegraphics[width=0.35\textwidth]{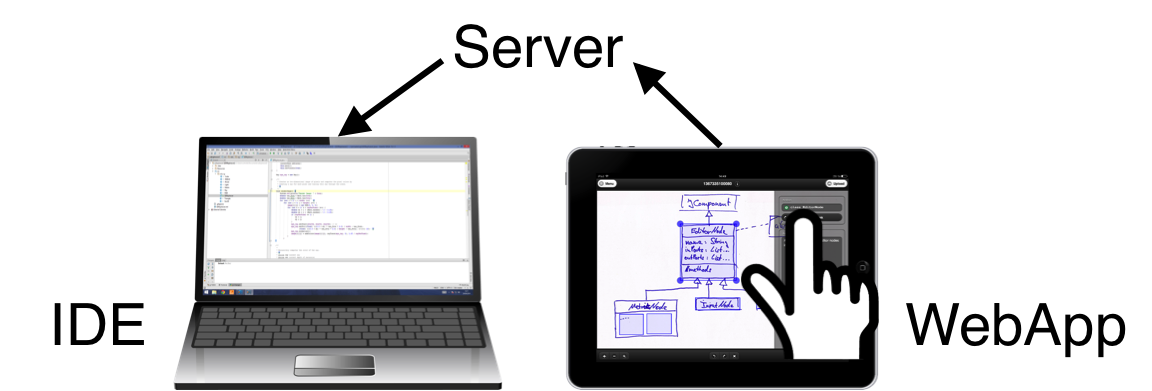}
\caption{Source code navigation using tablet}
\label{fig:navigation}
\end{figure}

\subsection{Link Anchors}

SketchLink uses a generic approach for linking sketches and diagrams to source code artifacts by employing universal \textit{link anchors}.
Every artifact that can be linked to other artifacts is identified by a \emph{Universally Unique Identifier} (UUID), to which a one-digit type identifier is prepended.
SketchLink currently supports three types of link anchors: \emph{source code anchors}, \emph{sketch anchors}, and \emph{marker anchors}.

\subsubsection{Source Code Anchors}

Source code anchors are either created by the server when the user links a sketch to a certain source code artifact using the web application or by the IDE plugin.
Our prototype currently only supports Java and embeds link anchors in Javadoc comments (see Figure \ref{fig:plugin}).
This has the advantage that links to the sketches on the server can be automatically inserted into the HTML documentation generated from Javadocs comments (at least for classes and methods).
If other statements or expressions are located in the same source code line as the comment, the anchor referes to this line.
Otherwise, the anchor refers to the subsequent element (e.g., a class or method decalaration).
Our approach is not limited to Java or Javadoc, because it only depends on the ability to insert an identifier in a source code comment, which is possible in every other programming language.

\subsubsection{Sketch and Marker Anchors}

The container format for sketch and marker anchors is SVG.
The sketch images are loaded using an SVG \texttt{image} element.
Users can link parts of a sketch to source code artifacts with rectangular markers (see Figure \ref{fig:workflow}c), which are stored as \texttt{rect} elements in the SVG.
The link anchors for a whole sketch or single markers are stored in the \texttt{id} attribute either of the SVG root element (sketch anchors) or of an rectangle element (marker anchors).

\subsection{Web Application}
\label{sec:webapp}

Using SketchLink on mobile devices like smartphones or tablets has the advantage that users can take a picture, for instance, of a whiteboard sketch, directly from the application running in a web browser, and upload it to the server (see Figure \ref{fig:workflow}).
When used in a desktop browser, the tool offers a file dialog for uploading image files like scanned sketches or digitally created diagrams.
After uploading the image file, users may add additional information like the authors of the sketch or a short description of the visual artifact.
This information is sent to the server, along with the image file of the sketch.
After the sketch is uploaded, the user can either link the entire sketch, or parts of it using a rectangular selection, to a source code artifact.
Furthermore, the user may annotate the selected areas with a text comment.
With support of our IDE plugin, the user can navigate to linked source code artifacts using the WebApp, either running on the same computer like the IDE or, for instance, a mobile device (see Figure \ref{fig:navigation}).
\subsection{IDE Plugin}
\label{sec:plugin}

When the plugin is started, it folds each UUID in the \verb+@sketchlink+ tags in Javadoc comments and hides them behind an icon (see Figure \ref{fig:plugin}).
If no other tags or comment text is present, Javadoc comments found inside a method are completely hidden.
The plugin was not only developed with the goal to explicitly visualize source code anchors, but to do this without distracting the developer during coding phases.
Therefore, the icon considers the current color scheme and automatically hides if its enclosing Javadoc comment gets folded.
Moreover, the icons---and the corresponding Javadoc tags---can be hidden globally.

When the user positions the mouse cursor over an icon, the linked source code is highlighted.
If the user clicks on an icon, a list with linked artifacts is shown (see Figure \ref{fig:plugin}a).
By positioning the mouse cursor over a list element, the user can open a preview of the linked sketch or marker (\textit{floating mode}).
Above the image, the authors of the sketch and its annotation are displayed (if this information is available on the server).
If markers are present and the user places the mouse cursor over one of them, the annotation for this marker is displayed instead of the annotation for the whole sketch and the linked source code is highlighted.
By left-clicking on a marker or a sketch, the user can navigate to the linked source code artifacts directly from within the preview window.
A right-click opens the configured web browser and loads the sketch in the SketchLink WebApp.
This way, the user may edit the annotation or create and link new markers.
The user can also switch to the \textit{docked mode} of the plugin, where a list of the linked sketches is displayed right next to the source code (see Figure \ref{fig:plugin}b).
Furthermore, the plugin assists the user in creating new source code anchors. After a new anchor is created, the plugin prompts the user to open the WebApp in order to directly link the newly created anchor to a sketch or marker.
For each source code anchor, metadata like the modification timestamp, the project name, the type of the linked artifact (e.g., class, method, if-statement), and the path to the source code file (relative to the project root) are stored on the server.

\section{Conclusion and Future Work}

Our first prototype named \emph{SketchLink} enables developers to easily capture, annotate, and link their sketches and diagrams to the related source code artifacts of Java projects using a web application and an IDE plugin.
The WebApp can also be used to navigate through a software project using created links.
Furthermore, our IDE plugin visualizes the links in situ in the source code editor and assists developers in creating new link anchors.

A future version of SketchLink should support the evolution of sketches and diagrams, because they are often reiterated and evolve over time \cite{Walny11-1, Cherubini07, Baltes14}.
However, not only the evolution of sketches has to be considered.
When software evolves, linked source code artifacts may be deleted or renamed.
Since the detection of moved or renamed source code artifacts between two revisions in the version control system is generally not an easy task \cite{Weissgerber06}, the plugin should consider common refactorings in order to keep the data on the server up to date.
Moreover, the tool could propose new links to the user by analyzing existing links and the relation between linked source code artifacts.
In the future, we want to enable users to retrieve captured sketches using metadata like the date, authors, or keywords found in the annotation.
Another future feature could be the visualization of the currently executed method in a linked sketch to support debugging tasks.
In addition, we plan to implement the possibility to link sketches or parts of sketches to other sketches.
One application of this feature would be the creation of a storyboard, allowing the user to ``play'' sketches in a defined order.
Moreover, linking markers with other sketches would be helpful to evaluate GUI mockups, because transitions from one view to another could be simulated.
Furthermore, it may be sensible to add a user management, allowing to differentiate between personal and shared sketches. 

Finally, we plan to execute both a user study evaluating the usability of our prototype and empirical studies to investigate if and how the availability of our tool supports the understanding of unfamiliar source code.
An interesting but hard-to-measure property is the value of captured and linked sketches.
It would be interesting to know how and when developers use the links.
We could investigate which properties valuable sketches possess and if visualizations for certain source code artifacts share common characteristics.

%
\bibliographystyle{abbrv}
\bibliography{bibliography}  
%
%
\end{document}